\documentclass[aps,prb,preprintnumbers,amsmath,amssymb,superscriptaddress,notitlepage]{revtex4-1}
\usepackage{graphicx}
\usepackage{dcolumn}
\usepackage{bm}
\usepackage[T1]{fontenc}
\usepackage[latin1]{inputenc}
\usepackage[english]{babel}
\usepackage{amsmath}
\usepackage{amssymb}
\usepackage{amstext}
\usepackage{braket}
\usepackage{relsize}
\usepackage{subfigure}
\usepackage{color}
\usepackage[rightcaption]{sidecap}
\usepackage{ulem}
\begin{document}
\bibliographystyle{unsrt}
\preprint{}
\title{Quantum simulation of frustrated magnetism in triangular optical lattices}
\author{J. Struck}
\affiliation{Institut f\"ur Laser-Physik, Universit\"at Hamburg, Luruper Chaussee 149, 22761 Hamburg, Germany}
\author{C. \"Olschl\"ager}
\affiliation{Institut f\"ur Laser-Physik, Universit\"at Hamburg, Luruper Chaussee 149, 22761 Hamburg, Germany}
\author{R. Le Targat}
\affiliation{Institut f\"ur Laser-Physik, Universit\"at Hamburg, Luruper Chaussee 149, 22761 Hamburg, Germany}
\author{P. Soltan-Panahi}
\affiliation{Institut f\"ur Laser-Physik, Universit\"at Hamburg, Luruper Chaussee 149, 22761 Hamburg, Germany}
\author{\\A. Eckardt}
\affiliation{
    ICFO -- Institut de Ci\`encies Fot\`oniques, Mediterranean Technology Park, E-08860 Castelldefels (Barcelona), Spain
    }
\author{M. Lewenstein}
\affiliation{
    ICFO -- Institut de Ci\`encies Fot\`oniques, Mediterranean Technology Park, E-08860 Castelldefels (Barcelona), Spain
    }
\affiliation{
    ICREA - Instituci\'o Catalana de Ricerca i Estudis Avancats, E-08010 Barcelona, Spain
    }
\author{P. Windpassinger}
\affiliation{Institut f\"ur Laser-Physik, Universit\"at Hamburg, Luruper Chaussee 149, 22761 Hamburg, Germany}
\author{K. Sengstock}
\email{sengstock@physik.uni-hamburg.de}
\affiliation{Institut f\"ur Laser-Physik, Universit\"at Hamburg, Luruper Chaussee 149, 22761 Hamburg, Germany}


\begin{abstract}
Magnetism plays a key role in modern technology as essential building block of many devices used in daily life. Rich future prospects connected to spintronics\cite{Zutic2004}, next generation storage devices\cite{Kimel2007} or superconductivity\cite{Lee2006} make it a highly dynamical field of research. Despite those ongoing efforts, the many-body dynamics of complex magnetism is far from being well understood on a fundamental level. Especially the study of geometrically frustrated configurations is challenging both theoretically and experimentally\cite{Balents2010}. Here we present the first realization of a large scale quantum simulator for magnetism including frustration. We use the motional degrees of freedom of atoms to comprehensively simulate a magnetic system in a triangular lattice. Via a specific modulation of the optical lattice, we can tune the couplings in different directions independently, even from ferromagnetic to antiferromagnetic. A major advantage of our approach is that standard Bose-Einstein-condensate temperatures are sufficient to observe magnetic phenomena like N\'eel order and spin frustration.  We are able to study a very rich phase diagram and even to observe spontaneous symmetry breaking caused by frustration.  In addition, the quantum states realized in our spin simulator are yet unobserved superfluid phases with non-trivial long-range order and staggered circulating plaquette currents, which break time reversal symmetry. These findings open the route towards highly debated phases like spin-liquids\cite{Balents2010,Sachdev2008} and the study of the dynamics of quantum phase transitions.
\end{abstract}
\maketitle
Frustrated spin systems, first studied by Pauling\cite{Pauling1935} and Wannier\cite{Wannier1950} more than 60 years ago, still belong to the most demanding problems of magnetism and condensed matter physics. The simplest realization of geometrical spin frustration is the triangular lattice as depicted in Figure 1 with antiferromagnetic interactions: the spins cannot order in the favored antiparallel fashion and have to compromise.  The rich variety of possible spin configurations\cite{Balents2010} arising from the competition between interactions and the geometry of the lattice has been studied in many different contexts. For example, spin-liquid phases in quantum magnetism are a very active field of research\cite{Balents2010}. Interestingly, also classical frustrated spin systems show intriguing properties\cite{Moessner2006,Morris2009,Fennel2009} like highly degenerate ground states, and emergent phenomena like artificial magnetic fields and monopoles observed in spin ice.

\begin{figure}
\includegraphics[width=0.49\textwidth]{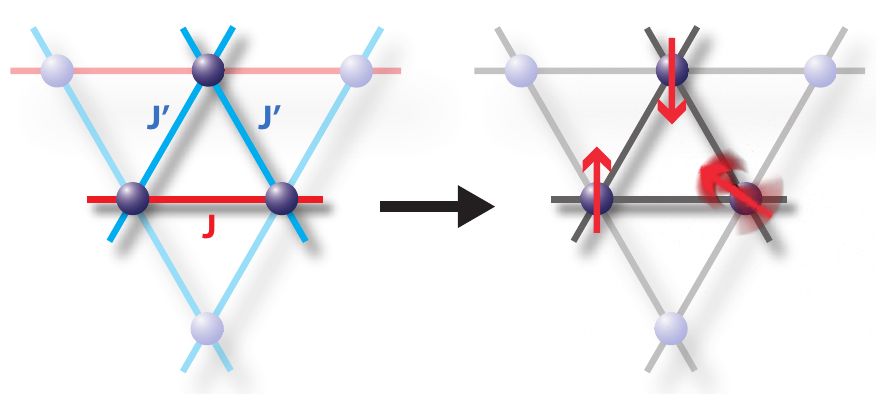}
        \caption{
       Illustration of a single plaquette within a large scale triangular lattice. The independent control parameters $J$ and $J'$ we access in the experiment are highlighted. The local phase of the atoms residing on a single lattice site is mapped onto a classical vector spin (red arrows). The coupling parameters $J$ and $J'$ (which can be tuned ferro- or anti\-ferromagnetic) determine the resulting spin configuration.
        }
        \label{fig:1}%
\end{figure}

Despite the long lasting interest in magnetically frustrated systems, it still poses a severe challenge to realize and characterize them experimentally in "natural" solid state devices. Here, the concept of a quantum simulator comes into play: to replace the complex, not very well controllable system with a well tunable quantum system which mimics the relevant physics. Recently, there have been considerable efforts in this direction and various schemes of simulating magnetism have been reported\cite{Lewenstein2007,Bloch2008}. Small systems of two quantum spins coupled by superexchange\cite{Trotzky2008a}, three Ising spins on a frustrated triangular plaquette\cite{Kim2010}, and four Heisenberg spins on a frustrated square plaquette\cite{Ma2011} were successfully simulated. Systems with large particle numbers, on the other hand, have been used to emulate artificial magnetic fields for atoms\cite{Lin2009}, itinerant ferromagnetism\cite{Jo2009} and very recently antiferromagnetic chains of Ising spins\cite{Simon2011}.

Here we report on a versatile simulator for large-scale magnetism on a two-dimensional triangular optical lattice\cite{Becker2010} by exploiting the motional degrees of freedom of ultracold bosons\cite{Eckardt2010}. The cornerstone of our simulation is the independent tuning of the nearest neighbor coupling elements $J$ and $J'$ (Figure 1) by introducing a fast oscillation of the lattice. In particular, we can even control their sign\cite{Eckardt2005,Lignier2007}, thus allowing for ferromagnetic or antiferromagnetic schemes at will. Thereby we gain access to the whole diversity of expected complex magnetic phases in our two dimensional triangular system and can study large system phase transitions as well as spontaneous symmetry breaking caused by frustration. With our approach, the rather easily achievable Bose-Einstein condensate (BEC) temperatures are sufficient to observe N\'eel-ordered and spin-frustrated states. This is an advantage compared to systems based on superexchange interaction which demand for much lower temperatures. In the following we will explain the main ideas of our quantum simulator of magnetism.

\begin{figure*}
\includegraphics[width=0.99\textwidth]{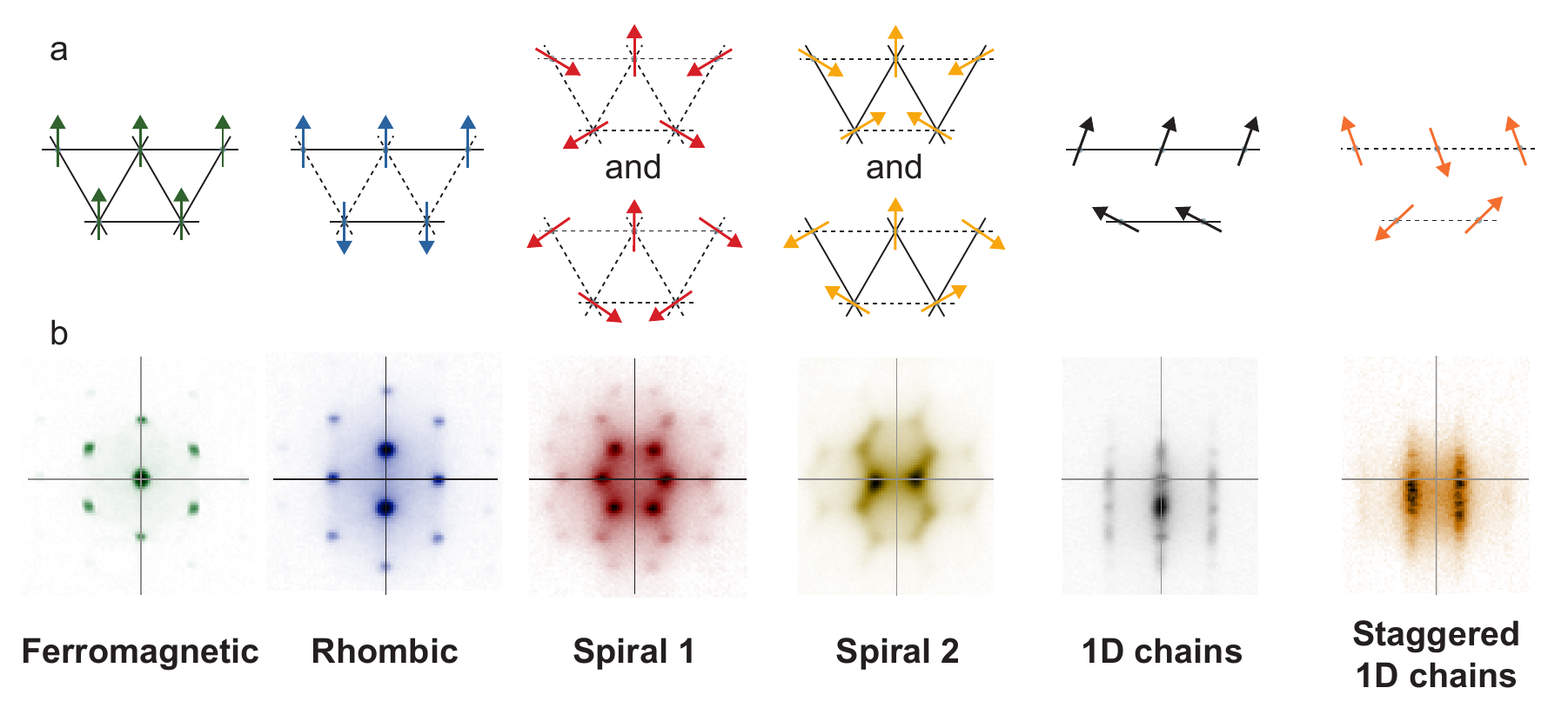}
    \caption{
    Spin configurations in a triangular lattice and their experimental signatures. a) Sketches of small parts of the six relevant spin-orders which can be realized within the large scale lattice by tuning $J$ and $J'$ are shown. Solid and dashed lines indicate ferromagnetic and antiferromagnetic couplings, respectively. In the spiral cases, two energetically degenerate spin configurations exist. b) Corresponding momentum distributions as experimentally observed. The axes in the experimental data mark the absolute position of the peaks. The pictures represent averages of several experimental runs. In the two spiral cases, since both ground state configurations randomly appear, the signature of both modes is present in the average of consecutive pictures (see Figure 4). For the 1D chains we observe a collection of the ferromagnetic and the rhombic structures.
    }
    \label{fig:2}%
\end{figure*}

For weak interactions ultracold bosonic atoms in an optical lattice form a superfluid state. In this case the atoms at each site i of the lattice have a well defined local phase $ \theta_i$ which can -- as a central concept here -- be identified with a classical vector spin $\mathbf{S}_i=(\cos(\theta_i) ,\sin(\theta_i ))$ (see also Figure 1). Long range order of these local phases (=spins) is imprinted by the minimization of the energy:
\begin{equation}
E({\theta_i })= -\sum_{\langle i,j \rangle}{J_{ij}  \cos(\theta_i-\theta_j ) }=-\sum_{\langle i,j \rangle}{J_{ij} \mathbf{S}_i \cdot \mathbf{S}_j }
\end{equation}
where the sum extends over all pairs of neighboring lattice sites. Note that we study large systems of about 10000 populated lattice sites. As a second central concept, the tunneling matrix elements $J_{ij}$ assume the role of the "spin-spin" coupling parameters between neighboring lattice sites -- positive $J_{ij}$ correspond to ferromagnetic interaction, negative $J_{ij}$ to antiferromagnetic. The most important feature of our approach is the independent tuning of the tunneling parameters $J$ and $J'$ along two directions (Figure 1) via an elliptical shaking of the lattice20\cite{Eckardt2010}.  This leads to various ferromagnetic, antiferromagnetic as well as mixed spin configurations as depicted in Figure 2a. In the situation where all tunneling parameters are positive $(J,J'>0)$, the spins align parallel and we associate this with a fully ferromagnetically ordered phase. This is identical to the ordering observed without shaking. When, for example, the sign of the $J'$-couplings are inverted $(J>0,J'<0)$, the new ground state of the system is of rhombic order: along the direction of negative coupling, the spins arrange in antiferromagnetic order while the third direction remains ferromagnetic. The other spin configurations depicted in Figure 2a -- spiral and chain order -- can be explained in a similar fashion. Each of these spin configurations has its own, unique quasi-momentum distribution, which serves as a clear signature for identification. The experimental data obtained for the different spin configurations are presented in Figure 2b.

The rich variety of spin orders as a function of the control parameters $J$ and $J'$ can be mapped into the phase diagram shown in Figure 3a. The background colors indicate the different spin configurations as expected from our calculations. For the data points, the color is chosen according to the spin configuration inferred from the experimental data. The measured data matches very well with theory. The phase diagram has several intriguing features: First, the ferromagnetic phase (F) on the right hand side $(J'>0)$ extends into regions where the J-coupling already favors antiferromagnetic order. The same behavior is observed for the rhombic phase (R) $(J'<0)$ which interestingly extends to regions $(J,J'<0)$ with purely antiferromagnetic couplings. Only for $J<-|J'|/2$, frustration finally breaks the (anti-)parallel spin order and leads to phases characterized by spiral spin configurations (S1, S2). In this region the system possesses two energetically degenerate ground states, which we will discuss in more detail below. Second, the transitions between the phases even are of different nature. The transition from ferromagnetic (F) to rhombic (R) is of first order.  In consequence, the indicated ferromagnetic chain order (C), as it is expected exactly at the phase boundary, represents an unstable situation which we cannot resolve experimentally. Instead we see a collection of interference peaks belonging to both neighboring phases (Figure 2b). The phase transitions into the spiral region (R to S1 and F to S2) are of second order. Finally, within the spiral region of the phase diagram, the spin configurations smoothly evolve. As a consequence, the staggered chain order (SC) found around $J'=0$ is stable and experimentally well observable (Figure 2b). The grey shaded region in the center indicates that for small values of $|J|$ and $|J'|$ the long range spin order is lost.

\begin{figure*}
\includegraphics[width=.99\textwidth]{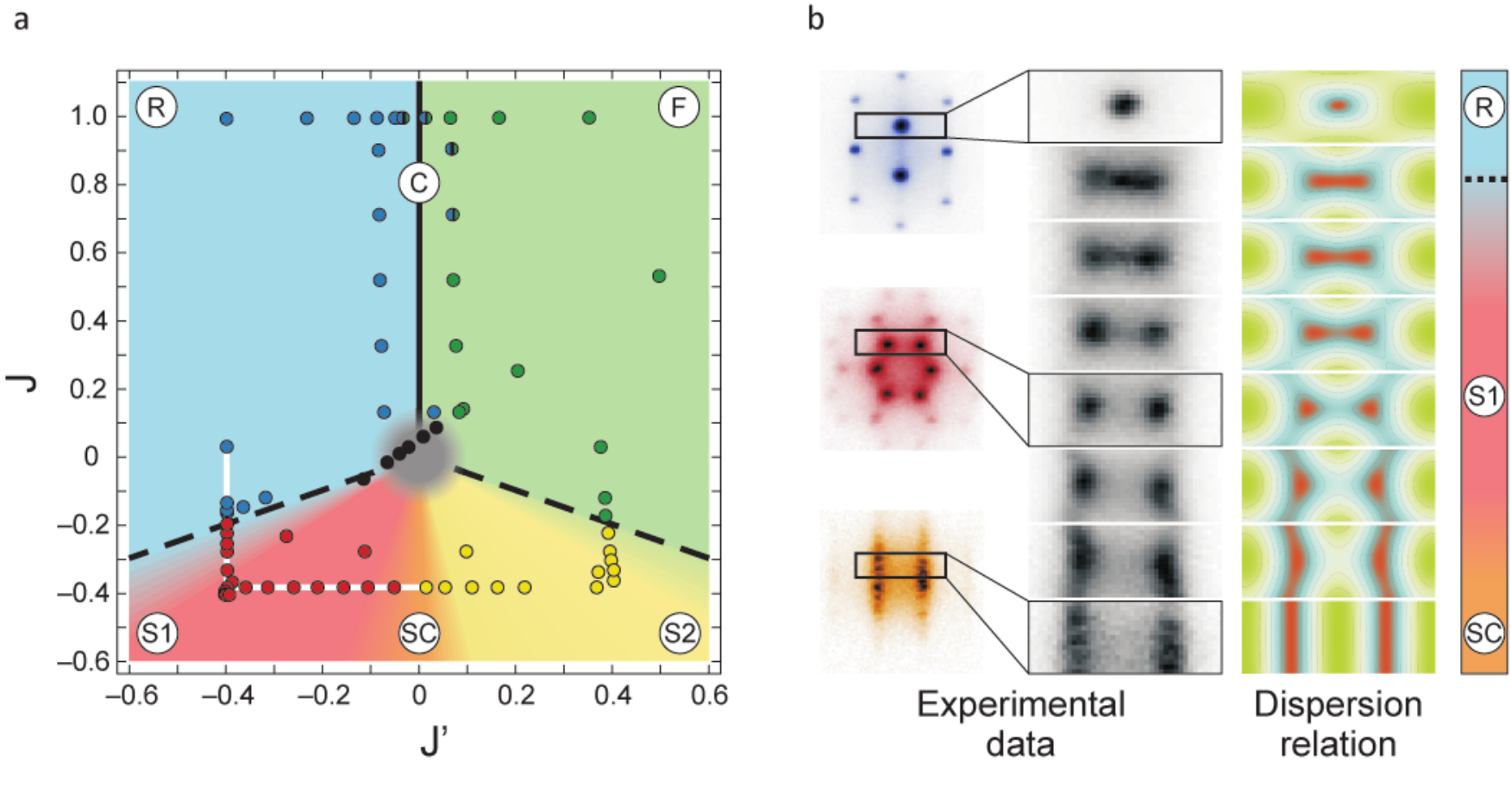}
        \caption{Mapping of the magnetic phase diagram. a) Phase diagram as a function of J and J', both given in units of the bare tunneling matrix elements.  J=J'=1 corresponds to no shaking of the lattice.  Letters label the spin configurations in the following way: F: Ferromagnetic, C: 1D chains, R: Rhombic, S1: Spiral 1, SC: staggered 1D chains and S2: Spiral 2. The black lines indicate the different phase transitions: the solid line marks a first-order transition, while dashed lines indicate second-order transitions. The color of the data points indicates the spin configuration inferred from the experimental data. b) Continuous change of the momentum structure in the spiral phase. On the left we plot three examples of the whole momentum peak structure we experimentally observe for the rhombic order, the isotropic spiral phase and the staggered chains. The zoomed in, uncolored data show the evolution of the relevant peaks for measurements along the white line indicated in a). For comparison we also display the calculated lowest band dispersion relation on the right hand side which indicate where we theoretically expect the momentum peaks (highest density in red).
        }
        \label{fig:3}%
\end{figure*}

In Figure 3b we study in detail the continuous phase transition from rhombic to spiral S1 order and the evolution of the state along a specific trajectory. We observe a single momentum peak in the rhombic region of the phase diagram that smoothly splits into two once the trajectory enters the spiral region. This is in full agreement with the calculated dispersion relation also shown in the figure. The development of these two peaks marks the existence of two degenerate ground states. In this situation, the system is expected to randomly choose one of the possible states, thus to exhibit spontaneous symmetry breaking. We consider this important feature in more detail for the particular case of isotropic antiferromagnetic couplings $(J=J'<0)$. The two possible emerging ground-state spin configurations shown in Figure 4a are mirror images of each other and can thus be distinguished by the chiral order parameter\cite{Diep2004} $({\kappa=\text{sgn}(\mathbf{S}_1 \times \mathbf{S}_2+\mathbf{S}_2 \times \mathbf{S}_3+\mathbf{S}_3 \times \mathbf{S}_1)}_z)$ of upwards and downwards pointing plaquettes. Their different momentum distributions (Figure 4b) allow for a direct distinction in the experiment. For consecutive experimental runs, each of the two spin configurations appears fully randomly (Figure 4c), which is a clear evidence for the spontaneous nature of the symmetry breaking between the two ground states. In most cases (Figure 4d) one of the modes clearly dominates, however, in few cases both configurations are found at the same time. The latter might be due to excitations like the formation of spin domains. In conclusion for this part, our system allows for the first time detailed studies on the mechanism of symmetry breaking in large scale "magnetic" systems.

\begin{figure*}
\includegraphics[width=0.99\textwidth]{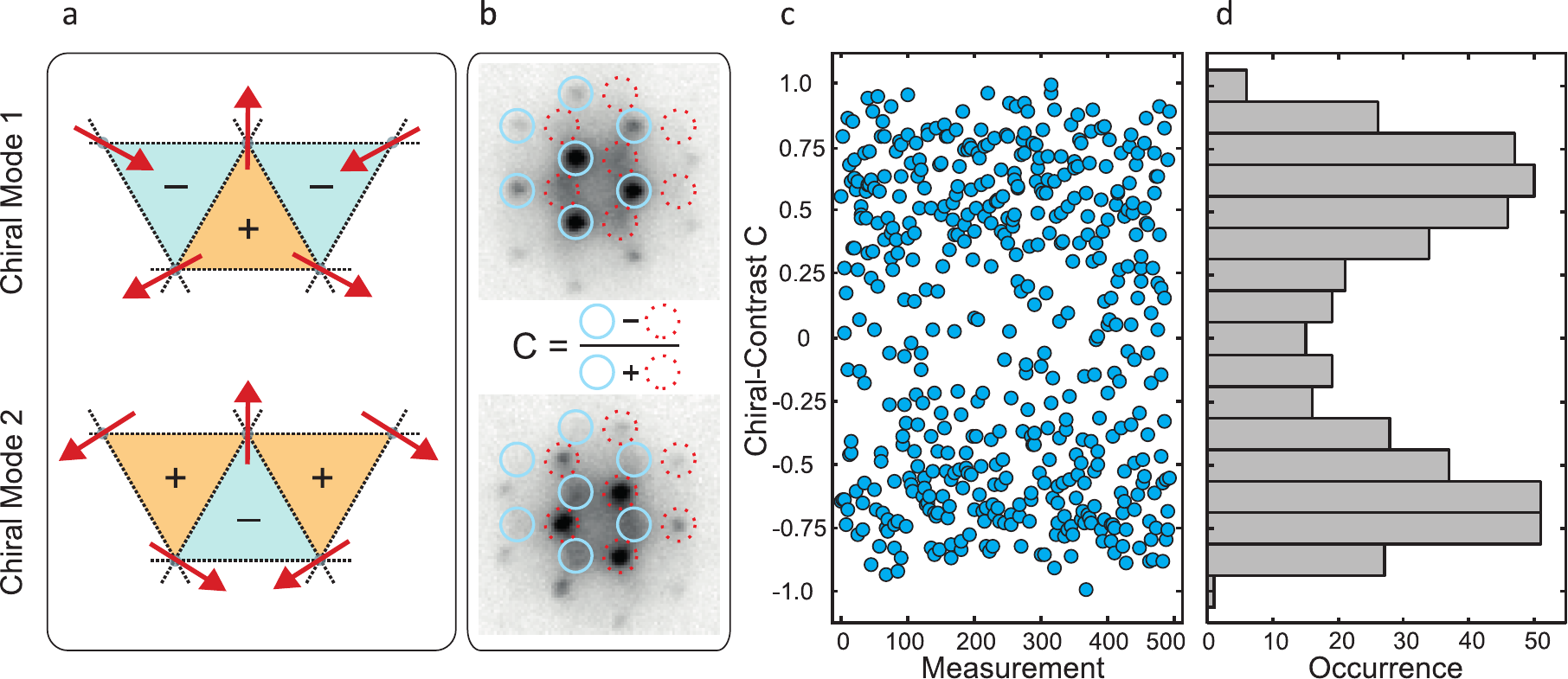}
    \caption{Spontaneous symmetry breaking in the spiral 1 region for J=J'=-0.4. a) Two different chiral configurations exist in the spiral case. The plus and minus signs indicate the value of the chiral order parameter. b) The distinct momentum distributions of the modes allow for their unambiguous identification. The chiral contrast C defined in the figure allows us to quantify the occupancy of the two different orders. Positive values indicate a higher occupancy of the chiral mode 1 while negative values indicate the opposite. c) Experimentally observed chiral contrast of around 500 consecutive measurements. The contrast is normalized to the highest absolute value of the series. d)  Corresponding histogram for the measurement series of c) with a binning width of 0,125.
    }
        \label{fig:4}%
\end{figure*}

A further fascinating aspect of our system is revealed when recalling that the discussed spin configurations actually correspond to local phases of a Bose-Einstein condensate at different sites of the triangular lattice. This provides new insight into unconventional superfluidity\cite{Wirth2011}: for all phases but the ferromagnetic one, the state corresponds to a superfluid at non-zero quasi-momentum and thus non-trivial long range phase order. Moreover, the observed spiral configurations spontaneously break time-reversal symmetry by showing circular bosonic currents around the triangular plaquettes of the lattice. Clock- and anticlockwise currents are found in a staggered fashion from plaquette to plaquette. These resemble the currents of the staggered flux state conjectured to play a role in explaining the pseudogap phase of high-temperature cuprate superconductors\cite{Lee2006}.

The obtained results demonstrate the realization of a quantum simulator for classical magnetism in a triangular lattice. Let us point out here, that this is performed by using simply spinless bosons. Due to the high degree of controllability we succeeded in observing all the various magnetic phases and phase transitions of first and second order as well as frustration induced spontaneous symmetry breaking. For the first time it becomes possible to quench systems on variable timescales from ferromagnetic to antiferromagnetic couplings and study the complex relaxation dynamics. Furthermore, extending the studies to the strongly correlated regime promises to give a deeper insight into the understanding of quantum spin models and spin liquid like phases\cite{Schmied2008}.

\acknowledgements{
We acknowledge stimulating discussions with A. Rosch, P. Hauke and D.-S. L\"uhmann. The work has been funded by DFG grants FOR 801, GRK 1355 and by the Landesexzellenzinitiative Hamburg which is supported by the Joachim Herz Stiftung. AE and ML are grateful for the support through the Spanish MINCIN  Grant  TOQATA, ERC Grant QUAGATUA, EU Grants AQUTE and NAMEQUAM.
}



\end{document}